\documentclass[twocolumn,showpacs,amsmath,amssymb]{revtex4}
\usepackage{graphicx,color}

\begin{document}

\title{Diffusion in a crowded environment}

\author{Duccio Fanelli$^{1}$ and Alan J. McKane$^{1,2}$}

\affiliation{$^{1}$Dipartimento di Energetica, University of Florence and INFN, 
Via S. Marta 3, 50139 Florence, Italy}

\affiliation{$^{2}$Theory Group, School of Physics and Astronomy, University of
Manchester, Manchester M13 9PL, U.K.}

\begin{abstract}
We analyze a pair of diffusion equations which are derived in the infinite 
system--size limit from a microscopic, individual--based, stochastic model. 
Deviations from the conventional Fickian picture are found which ultimately 
relate to the depletion of resources on which the particles rely. The 
macroscopic equations are studied both analytically and numerically, and are 
shown to yield anomalous diffusion which does not follow a power law with 
time, as is frequently assumed when fitting data for such phenomena. These 
anomalies are here understood within a consistent dynamical picture which 
applies to a wide range of physical and biological systems, underlining the 
need for clearly defined mechanisms which are systematically analyzed to 
give definite predictions. 
\end{abstract} 

\pacs{05.60.Cd, 05.40.-a, 87.15.Vv} 

\maketitle

Almost all discussions relating to the modeling of diffusion assume Fick's 
law --- that the rate at which one substance diffuses through another is 
directly proportional to the concentration gradient of the diffusing 
substance~\cite{diffBook}. There is a good reason for this: it is empirically 
very well supported, at least at not too high concentrations, and is part
of the wider theoretical framework of linear nonequilibrium 
thermodynamics~\cite{deG84}. However one might expect nonlinear corrections in
various situations e.g.~if obstacles are present or if there is more than one 
species at high concentrations, and indeed many reports of anomalous behavior 
can be found in the literature~\cite{bou90,metz00,Zas02,Fed96,Sh97,Bro99,Sax01,weiss03,Gui08}. These studies are mainly experimental, or involve numerical 
simulations, and the theory which is given mainly consists of phenomenological 
fits to the data. Many of the fits suggest that the mean-square displacement of
the diffusing species grows with time, $t$, like $t^{\alpha}$, where 
$\alpha < 1$. The phenomena described go under various 
names such as ``molecular crowding''~\cite{weiss04,banks,dix08} and 
``single-file diffusion''~\cite{Par94,Ke00,Wei00,brz07} , all indicating that 
the increased concentration impedes the flow of particles in some way, 
leading to a violation of Fick's law. 

There are relatively few first-principles studies of these effects. Those that 
do exist include hydrodynamical models with nonlinear constitutive relations 
~\cite{adk63}  and simple symmetric exclusion processes in one 
dimension~\cite{brz07}. In this Letter we propose a modification of Fick's law 
which is based on a physically motivated microscopic theory. Unlike previous 
theories of single-file diffusion~\cite{Har65,Per74}  it holds in arbitrary 
dimensions, since it is not a consequence of the physical `jamming' of the 
particles, rather it is due to the depletion of resources on which the 
particles rely. This resource may be space to move, but could also be a 
chemical substrate required for the system to remain viable. The phenomenon we
describe is seen in experiments which trace the motion of tagged particles.
This is equivalent to assuming the existence of two types of particle with the
same diffusion constant.

We begin with a generic microscopic system in a given volume of $d-$dimensional
space which is divided into a large number, $\Omega$, of small (hypercubic) 
patches. Each patch, labeled by $\alpha$, can contain up to $N$ particles: 
$n_{\alpha}$ of type $A$, $m_{\alpha}$ of type $B$, and 
$v_{\alpha} = N-n_{\alpha}-m_{\alpha}$ vacancies, denoted by $E$. We will assume 
that the particles have no direct interaction, however there will be an 
indirect interaction in that the mobility of particles will be affected if 
neighboring patches have few vacancies.

To model this more concretely we assume that the particles move only to 
nearest-neighbor patches, and then only if there is a vacancy there:
\begin{eqnarray}
& & A_{\alpha} + E_{\beta} \stackrel{\mu_{1}}{\longrightarrow} E_{\alpha} + 
A_{\beta}, \nonumber \\
& & B_{\alpha} + E_{\beta} \stackrel{\mu_{2}}{\longrightarrow} E_{\alpha} + 
B_{\beta}.
\label{mig}
\end{eqnarray}
Here $\alpha$ and $\beta$ label nearest-neighbor patches with 
$A_{\alpha}, B_{\alpha}$, and $E_{\alpha}$ being the respective types of particles
in patch $\alpha$, and $\mu_1$ and $\mu_2$ being the reaction rates. The state 
of the system will be characterized by the number of $A$ and $B$ particles in 
each patch, that is, by the vector
$\textbf{n}=(\textbf{n}_{1},\ldots,\textbf{n}_{\Omega})$, where 
$\textbf{n}_{\alpha}=(n_{\alpha},m_{\alpha})$. The rate of transition from state
$\textbf{n}'$, to another state $\textbf{n}$, is denoted by
$T(\textbf{n}|\textbf{n}')$ --- with the initial state being on the right.
The transition rates associated with the migration between nearest-neighbor
patches take the form
\begin{eqnarray}
T(n_{\alpha}-1,n_{\beta}+1|n_{\alpha},n_{\beta}) &=&
\frac{\mu_1}{z\Omega}\frac{n_{\alpha}}{N} \frac{N-n_{\beta}-m_{\beta}}{N}, 
\nonumber \\
T(m_{\alpha}-1,m_{\beta}+1|m_{\alpha},m_{\beta}) &=&
\frac{\mu_2}{z\Omega}\frac{m_{\alpha}}{N} \frac{N-n_{\beta}-m_{\beta}}{N},
\nonumber \\
\label{TRs}
\end{eqnarray}
where $z$ is the number of nearest neighbors that each patch has and where 
within the brackets we have chosen to indicate only the dependence on those 
particles which are involved in the reaction. It is the presence of the factor
$v_{\beta} = N-n_{\beta}-m_{\beta}$, which reduces the transition rate if there
are few vacancies in the target patch, and which modifies Fick's law in the 
macroscopic theory.

This is a Markov process, and the probability of finding the system in state 
$\textbf{n}$ at time $t$, denoted by $P(\textbf{n},t)$, is given by the 
master equation
\begin{equation}
\frac{dP(\textbf{n},t)}{dt} = \sum_{\textbf{n}'\neq\textbf{n}} 
\left[ T(\textbf{n}|\textbf{n}')P(\textbf{n}',t) 
- T(\textbf{n}'|\textbf{n})P(\textbf{n},t) \right],
\label{master}
\end{equation}
where the allowed transitions are those given by Eq.~(\ref{mig}). This defines 
the microscopic process, but we are interested in the macroscopic equations 
that this process generates. To find these we need to find the dynamical 
equations for the ensemble averages 
$\langle n_{\alpha} \rangle$ and $\langle m_{\alpha} \rangle$. Multiplying 
Eq.~(\ref{master}) by $n_{\alpha}$ and summing over all $\textbf{n}$ gives,
after shifting some of the sums by $\pm 1$,
\begin{eqnarray}
\frac{d\langle n_{\alpha} \rangle}{dt} &=& \sum_{\beta \in \alpha} 
\left[ \langle T(n_{\alpha}+1,n_{\beta}-1|n_{\alpha},n_{\beta}) \rangle 
\right. \nonumber \\
&-& \left. \langle T(n_{\alpha}-1,n_{\beta}+1|n_{\alpha},n_{\beta}) 
\rangle \right], 
\label{pre_macro}
\end{eqnarray}
where the notation $\sum_{\beta \in \alpha}$ means `sum over all patches $\beta$
which are nearest-neighbors of the patch $\alpha$'. A similar equation holds
for $d{\langle m_{\alpha} \rangle}/dt$.

The averages in Eq.~(\ref{pre_macro}) are carried out by using the explicit 
forms (\ref{TRs}) and replacing the averages of products by the products of 
averages, which is valid in the limit $N \to \infty$. Then scaling time by
a factor of $N\Omega$ one finds~\cite{mck04}
\begin{eqnarray}
\frac{d\phi_{\alpha}}{dt} &=& \mu_1 \left[ \Delta \phi_{\alpha} 
+ \phi_{\alpha} \Delta \psi_{\alpha} - \psi_{\alpha} \Delta \phi_{\alpha} \right],
\nonumber \\
\frac{d\psi_{\alpha}}{dt} &=& \mu_2 \left[ \Delta \psi_{\alpha} 
+ \psi_{\alpha} \Delta \phi_{\alpha} - \phi_{\alpha} \Delta \psi_{\alpha} \right].
\nonumber \\
\label{macro}
\end{eqnarray}
Here
\begin{equation}
\phi_{\alpha} = \lim_{N \to \infty} \frac{\langle n_{\alpha} \rangle}{N}; \ \ 
\psi_{\alpha} = \lim_{N \to \infty} \frac{\langle m_{\alpha} \rangle}{N},
\label{phi_psi_def}
\end{equation}
and $\Delta$ is the discrete Laplacian operator defined by
$\Delta f_{\alpha} = (2/z) \sum_{\beta \in \alpha} (f_{\beta} - f_{\alpha})$. 
Finally taking the size of the patches to zero, and scaling the rates $\mu_1$
and $\mu_2$ appropriately~\cite{mck04} to give diffusion constants $D_1$ and
$D_2$, gives partial differential equations
for $\phi(\textbf{x},t)$ and $\psi(\textbf{x},t)$:
\begin{eqnarray}
\frac{\partial \phi}{\partial t} &=& D_1\left[ \nabla^2 \phi 
+ \phi \nabla^2 \psi - \psi \nabla^2 \phi \right], \nonumber \\
\frac{\partial \psi}{\partial t} &=& D_2 \left[ \nabla^2 \psi
+ \psi \nabla^2 \phi - \phi \nabla^2 \psi \right],
\label{pdes}
\end{eqnarray}
where $\nabla^2$ is the usual Laplacian.

We can give a quite complete analysis of Eqs.~(\ref{pdes}) in the case when the
diffusion constants are equal. Let $D_{1}=D_{2}\equiv D$ and absorb $D$ into
the definition of the time. Then adding the two equations gives
\begin{equation}
\frac{\partial \rho}{\partial t} = \nabla^2 \rho, \ \ \rho \equiv 
\frac{1}{\sqrt{2}} \left( \phi + \psi \right),
\label{def_rho}
\end{equation}
whereas the equation for the difference $\sigma \equiv (\phi-\psi)/\sqrt{2}$ is
\begin{equation}
\frac{\partial \sigma}{\partial t} = \nabla^2 \sigma + \sqrt{2}
\left( \sigma \nabla^2 \rho - \rho \nabla^2 \sigma \right).
\label{sigma_real}
\end{equation}
We will take initial conditions such that $\rho(-\textbf{x},0)=\rho(\textbf{x},0)$ and
$\sigma(-\textbf{x},0)=-\sigma(\textbf{x},0)$. 
Solving Eq.~(\ref{def_rho}) for $\rho$ and going over to Fourier space gives
for Eq.~(\ref{sigma_real}):
\begin{eqnarray}
& & \frac{\partial \sigma(\textbf{k},t)}{\partial t} = 
-k^{2}\sigma(\textbf{k},t) + \sqrt{2} \int \frac{d\textbf{p}}{(2\pi)^d}\, 
\left[ p^2 - (\textbf{k}-\textbf{p})^2 \right] \nonumber \\
& & \times  \rho(\textbf{k}-\textbf{p},0)
e^{-(\textbf{k}-\textbf{p})^{2} t}\,\sigma(\textbf{p},t),
\label{sigma_fourier}
\end{eqnarray}
where $\rho(\textbf{k}-\textbf{p},0)$ is the initial value of $\rho$ in Fourier space. 
This equation is linear in $\sigma$ which allows us to make further analytic progress. We have 
considered two types of particle for simplicity; the above discussion can 
easily be extended to three or more types.

We will first analyze Eq.~(\ref{sigma_fourier}) in one dimension. The 
calculation in $d-$dimensions is not much more difficult, and we will give the 
main result later, but is less clear due to the number of indices involved in 
the intermediate steps. We begin by noting that $\rho(k,t)$ is even in $k$ and
$\sigma(k,t)$ is odd in $k$, so that $\sigma(0,t)=0$. So the first non-trivial
term in the expansion of $\sigma(k,t)$ is 
$\langle x(t) \rangle_{\sigma} \equiv -i \partial \sigma(k,t)\/\partial k|_{k=0}$. From Eq.~(\ref{sigma_fourier}) one finds that
\begin{equation}
i \frac{d\langle x(t) \rangle_{\sigma}}{dt} = 2 \sqrt{2} \int \frac{dp}{2\pi}\,p
\rho(p,0) e^{-p^{2}t}\,\sigma(p,t).
\label{eqn:mean}
\end{equation}
In order to evaluate the integral we need to know something of the behavior of
$\sigma(p,t)$. Numerical simulations (see later) indicate that the ratio of 
$\langle x^{2n+1}(t) \rangle_{\sigma}$ to 
$\langle x^{2n-1}(t) \rangle_{\sigma}$, $n=1,2,\ldots$, is proportional to $t$, 
for large $t$, to a very good approximation. Therefore
\begin{equation}
\sigma(p,t) = \sum^{\infty}_{n=0} 
\frac{(ip)^{2n+1} \langle x^{2n+1}(t) \rangle_{\sigma}}{(2n+1)!} \sim
i \langle x(t) \rangle_{\sigma} p f(p^{2}t),
\label{sigma_expansion}
\end{equation}
for large $t$ where
\begin{equation}
f(y) = \sum^{\infty}_{n=0} \frac{(-1)^{n} a_n y^{n}}{(2n+1)!},
\label{defn_f}
\end{equation}
and where the $a_n$ are constants. Substituting Eq.~(\ref{sigma_expansion}) 
into Eq.~(\ref{eqn:mean}) and scaling $p^2$ by $t$ gives
\begin{equation}
\frac{d\langle x(t) \rangle_{\sigma}}{dt} = \frac{A_1}{2 t^{3/2}}
\langle x(t) \rangle_{\sigma},
\label{diffeqn_mean}
\end{equation}
where
\begin{equation}
A_1 = \frac{4 V_1}{\pi} \int dp\,f(p^2) p^2 e^{-p^2}.
\label{defn_A}
\end{equation}
This scaling implies that $\rho(p,0)$ is replaced by 
$\rho(p=0,0)=\int dx \rho(x,0) \equiv \sqrt{2} V_1$, for large $t$.
Solving the differential equation (\ref{diffeqn_mean}) gives
\begin{equation}
\langle x(t) \rangle_{\sigma} = c\exp{\left(-\frac{A_1}{\sqrt{t}}\right)},
\label{result_mean_1}
\end{equation}
where $c$ is a constant.

The coefficients $a_n$ in Eq.~(\ref{defn_f}) may be found by differentiating 
Eq.~(\ref{sigma_fourier}) $(2n+1)$ times with respect to $k$ and then setting 
$k=0$. In the resulting differential equation for 
$\langle x^{2n+1}(t) \rangle_{\sigma}$ use of Eq.~(\ref{sigma_expansion}) 
shows that the $p-$integral is down by powers of $t$ on the contribution
coming from the term $-k^2 \sigma(k,t)$ for $n>0$. So for large $t$,
\begin{eqnarray}
\frac{d\langle x^{2n+1}(t) \rangle_{\sigma}}{dt} &\sim& 2n(2n+1) 
\langle x^{2n-1}(t) \rangle_{\sigma} \nonumber \\
&\sim& 2n(2n+1)a_{n-1} t^{n-1} \langle x(t) \rangle_{\sigma}.
\label{diffeqn_gen_n}
\end{eqnarray}
Using Eq.~(\ref{result_mean_1}) the integration in Eq.~(\ref{diffeqn_gen_n})
can be carried out for large $t$. This gives $\langle x^{2n+1}(t)\rangle_{\sigma}
\sim 2(2n+1)a_{n-1}t^{n}\langle x(t) \rangle_{\sigma}$ and so 
$a_{n}=2(2n+1)a_{n-1}$. Therefore from Eq.~(\ref{defn_f}) $f(y)=e^{-y}$ and
from Eq.~(\ref{defn_A}) $A_{1}=V_1/\sqrt{2\pi}$. Finally, from 
Eqs.~(\ref{sigma_expansion}) and (\ref{result_mean_1}) we find that
\begin{equation}
\sigma(k,t) \sim ikc \exp{\left(-\frac{V_1}{\sqrt{2\pi t}}\right)} 
\exp{\left(-k^2 t\right)},
\label{sigma_k}
\end{equation}
for large $t$. Taking the inverse Fourier transform gives
\begin{equation}
\sigma(x,t) \sim \frac{cx}{4 \pi^{1/2} t^{3/2}} 
\exp{\left( - \frac{V_1}{\sqrt{2\pi t}}\right)} \exp{\left( - x^{2}/4t \right)}.
\label{sigma_x_1}
\end{equation}

The corresponding calculation in $d-$dimensions can in principle be carried out 
in a similar way. This will be discussed in more detail elsewhere and here we 
only give the generalization of Eq.~(\ref{result_mean_1}):
\begin{equation}
\langle x_{i}(t) \rangle_{\sigma} = c_{i}\exp{\left(-\frac{A_d}{t^{d/2}}\right)},
\label{result_mean_d}
\end{equation}
where the $c_i$, $i=1,\dots,d$, are constants and $A_d$ is given by
\begin{equation}
A_d = \frac{8 V_d}{d^2}\int \frac{d\textbf{p}}{(2\pi)^d}\,p^2 e^{-2 p^2} = 
\frac{V_d}{2^{(3d-2)/2} \pi^{d/2} d}.
\label{defn_A_d}
\end{equation}

Returning to the one-dimensional case, we may use the results we have obtained
to characterize the time evolution of the variance of the original 
distributions $\phi$ and $\psi$. Since $\rho$ has only even moments and 
$\sigma$ only odd moments,
$\langle x^n \rangle_{\phi} = (-1)^n \langle x^n \rangle_{\psi}$, for any 
integer $n$. It follows that  
$\langle \langle x^2 \rangle \rangle_{\phi}=\langle \langle x^2 \rangle \rangle_{\psi}$, where the symbol $\langle \langle \cdot \rangle \rangle_f$ stands for 
the variance of function $f$. Using the result (\ref{result_mean_1}) for the 
first moment and the standard diffusion result for the second moment, we have for the normalized
mean--square displacement: 
\begin{equation}
\langle \langle x^2(t) \rangle \rangle_{\phi,\psi} = 2 t - \frac{c^2}{2 V_1} 
\exp{ \left[- V_1 \left(\frac{2}{\pi t}\right)^{1/2} \right]} + \frac{\langle x^2(0) \rangle_{\rho}}{\sqrt{2} V_1}.
\label{result_variance_1}
\end{equation} 
For large enough times, the system displays normal diffusion: the variances 
scale linearly with time, and are shifted by a constant factor. 
For relatively short times, of the order of the 
inverse of the diffusion coefficient, here absorbed in the definition of $t$, 
deviations from the usual behavior are predicted to occur due to the 
exponential factor in Eq.~(\ref{result_variance_1}), which reduces the 
diffusion. 

\begin{figure}[t!]
\centering
\vspace*{2.5em}
\includegraphics[width=6.5cm]{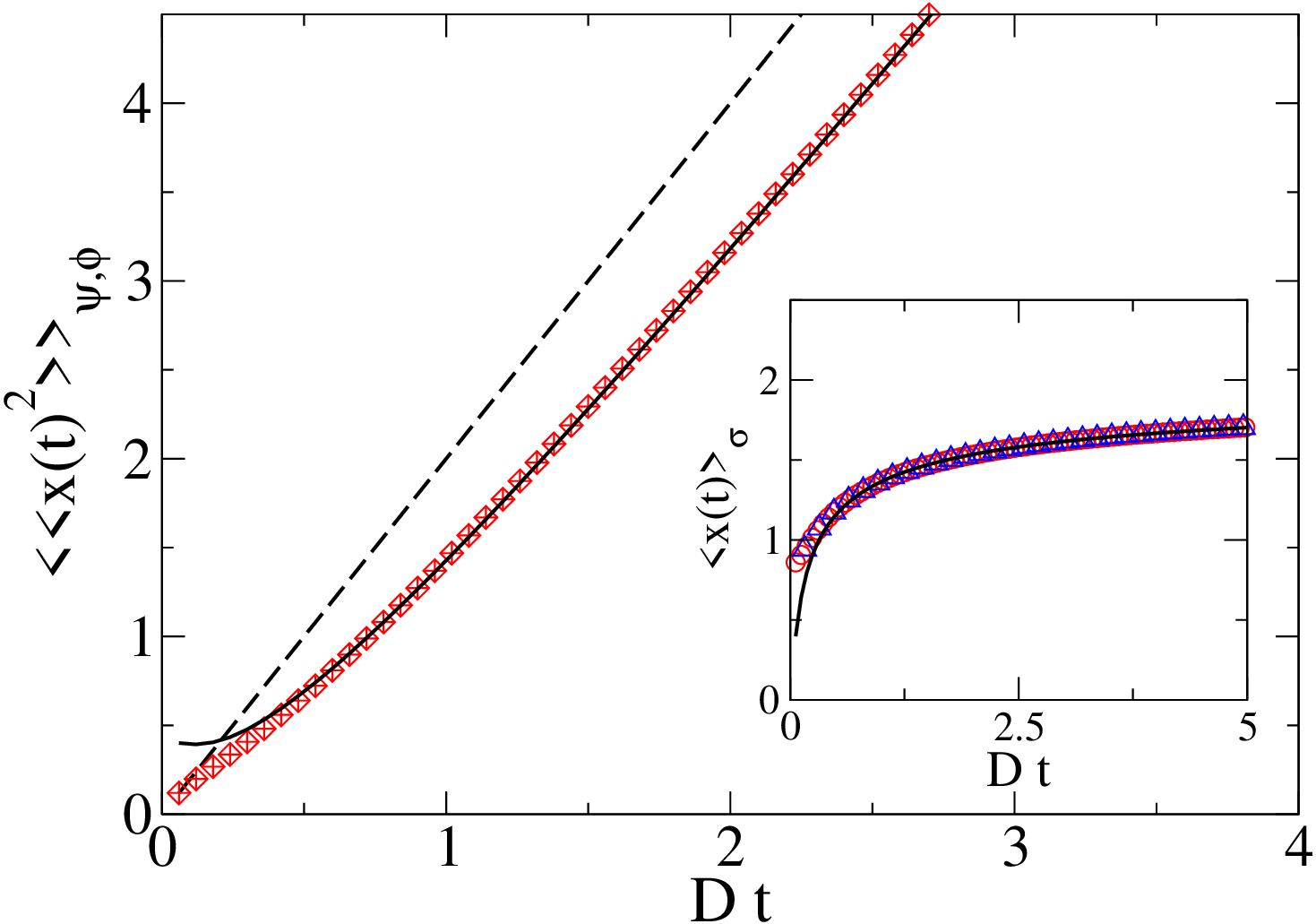}
\caption{(Color online) Main panel: the variances 
$\langle\langle x \rangle \rangle_{\phi}$ and 
$\langle\langle x \rangle \rangle_{\psi}$ are plotted as a function of the 
rescaled time, $D t$. Symbols refer 
to the numerical simulations, diamonds and pluses stand respectively for 
$\phi$ and $\psi$. The dashed line represents the normal diffusion prediction, 
while the thick solid line refers to formula (\ref{result_variance_1}) where 
$c=2.035$. Here $D=0.003$ and $V_1=1$.  Inset: The first moment 
$\langle x \rangle_{\sigma}$ is represented as a function of rescaled time, 
$D t$. Symbols refer respectively to $D=0.003$ (circles) and $D=0.008$ 
(triangles). The solid line is the asymptotic prediction (\ref{result_mean_1}) 
with $c=2.035$.}
\label{fig1}
\end{figure}

\begin{figure}[b!]
\centering
\vspace*{2.5em}
\includegraphics[width=6.5cm]{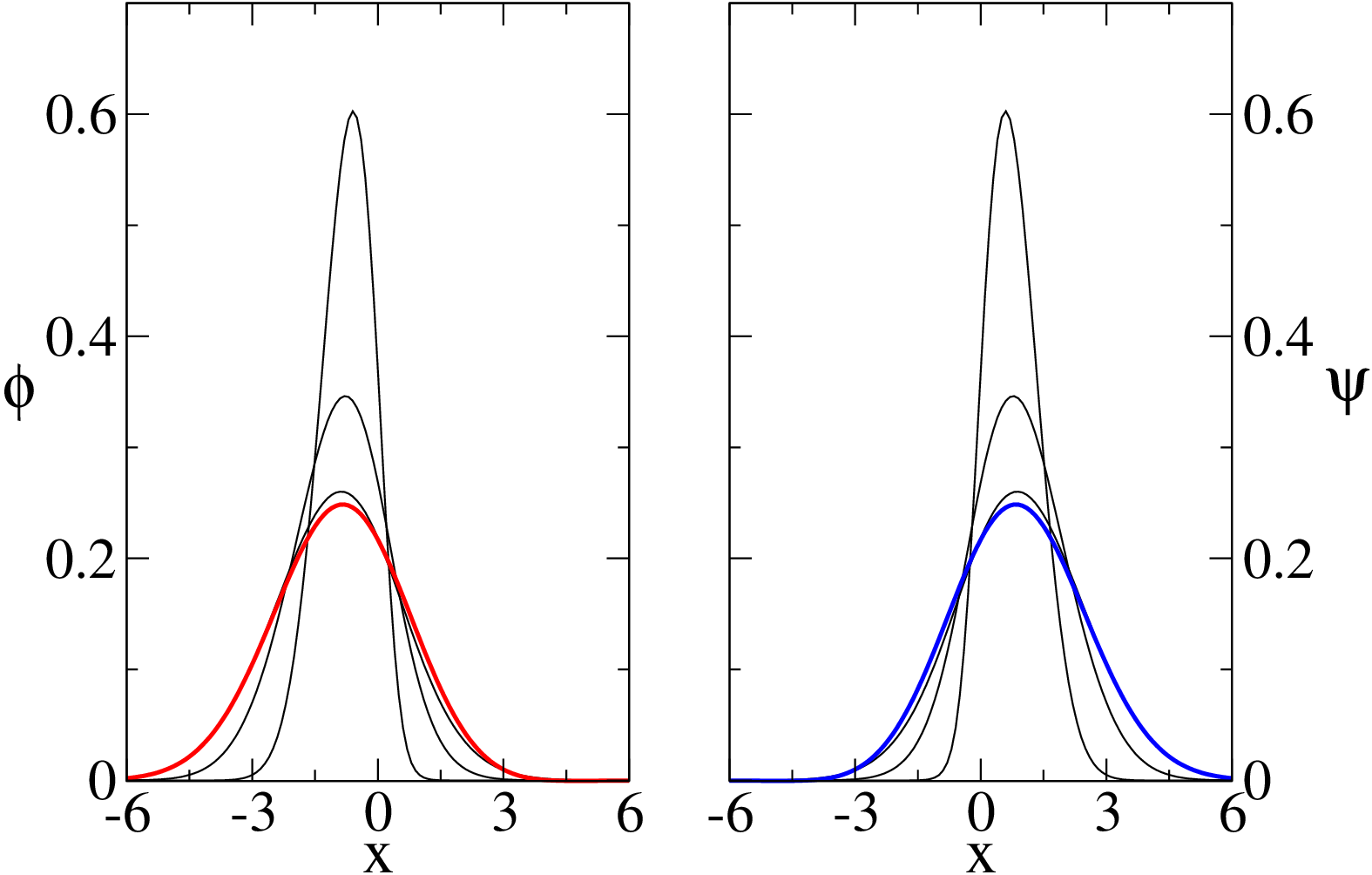}
\caption{(Color online) Left panel: three profiles of the distribution $\phi$ 
are plotted (thin line), corresponding to successive instants of 
the dynamics. From top to bottom, the snapshots are taken at rescaled time 
$Dt=0.3,0.9,1.5$. The theoretical profile at $Dt=1.5$ is represented with a 
thick line (red online). Right panel: Same as in left panel, but for the 
distribution $\psi$. The analytical prediction is plotted with a thick line 
(blue online). The simulations refer to $D=0.003$ and  $V_1=1$.}
\label{fig2}
\end{figure}

The validity of the approximations we have made in the above analysis have 
been checked by carrying out numerical simulations for the one-dimensional case.
We used Euler discretization, both in the space and time coordinates of 
Eq.~(\ref{pdes}) and assumed identical diffusion constants for both species, 
so as to make contact with the analysis. In the main panel of Fig.~\ref{fig1} 
the time evolution of the measured variances, for both $\phi$ and $\psi$, are 
displayed. The variances are seen to grow more slowly than predicted from
standard diffusion theory. The agreement between the theoretical prediction 
(\ref{result_variance_1}) and the simulations is excellent, even at quite 
short times. The constant $c$ is determined by fitting the late time evolution 
of $\langle x \rangle_{\sigma}$ from simulations to the asymptotic profile 
(\ref{result_mean_1}), see inset of Fig.~\ref{fig1}. The form of the time 
evolution of the first moment $\langle x \rangle_{\sigma}$ given by 
Eq.~(\ref{result_mean_1}) has also been checked numerically for a range of 
parameters, and is always in excellent agreement. As a further check, 
Fig.~\ref{fig2} shows $\phi$ and $\psi$ found from numerical simulations for 
a range of times. The recorded snapshots at the largest time show good 
agreement with the theoretical curves at that time.
 
\begin{figure}[t!]
\centering
\vspace*{2.5em}
\includegraphics[width=6.5cm]{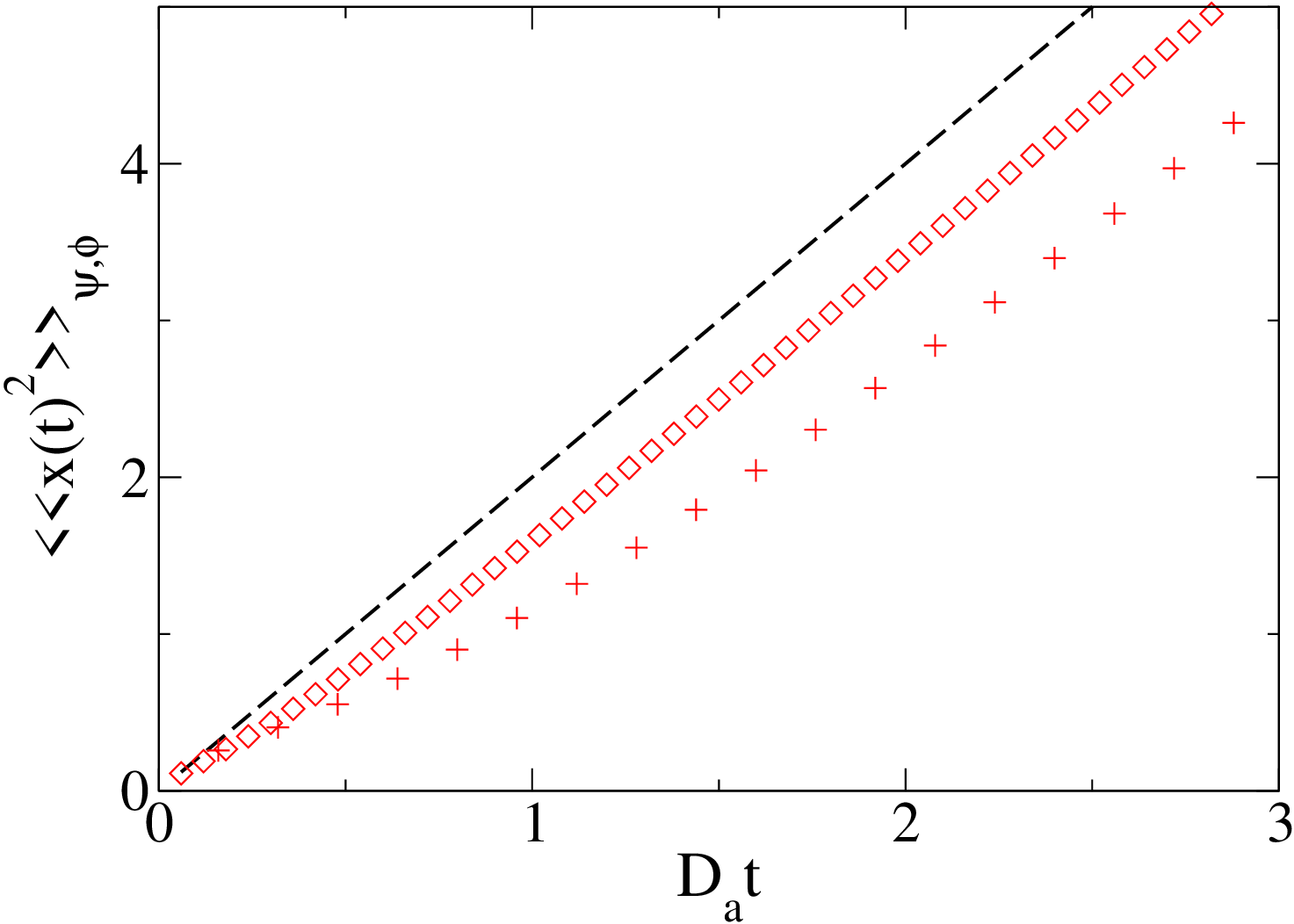}
\caption{(Color online) The variances  $\langle\langle x \rangle\rangle_{\phi}$
(diamonds) and $\langle\langle x \rangle \rangle_{\psi}$ (pluses) are plotted 
as a function of their respective rescaled time, $D_{a} t$, with 
$a=1,2$. In both cases the variances grow slower than predicted by the 
standard diffusion theory (dashed line). Here, $D_1=0.003$, $D_2=0.008$ and 
$V_1=1$.}
\label{fig3}
\end{figure}

These considerations all point to the validity of the analysis for the case
when the diffusivities of the two species are equal. To shed light on
the more general setting where $D_1 \ne D_2$, and so broaden the range of 
applicability of our conclusions, we rely on numerical simulation. In 
Fig.~\ref{fig3}, we show the time evolution of the variance of $\phi$ (resp. 
$\psi$) vs. the rescaled time $D_1 t$ (resp. $D_2 t$). As in the symmetric 
case, the growth of the variances is slower than for standard diffusion. This 
again reflects the presence of the finite carrying capacity imposed at the 
level of the microscopic dynamics. Remarkably, the function 
$\langle x \rangle_{\sigma}$ is again found to empirically obey 
Eq.~(\ref{result_mean_1}), the undetermined factor $A_1$ depending on the 
individual diffusivities. 

The modified diffusive behavior we have found is derived from a general 
principle formulated at the microscopic level, not from a phenomenological fit. 
It is important to stress this fact, since virtually all the work on molecular 
crowding, and related phenomena, to date has postulated that the mean-square
displacement increased in time like a power: 
$\langle \langle x^2 \rangle \rangle \sim t^{\alpha}$. There has been 
much discussion of whether $\alpha < 1$ (called `subdiffusion') or $\alpha > 1$
(called `superdiffusion'). Our results can be fitted by either, depending on 
when the time-window for the fit is taken; if taken at early times 
subdiffusion is found, at late times superdiffusion is found. Indeed by 
picking suitable time-windows a wide range of values of $\alpha$ can be found. 
This ambiguity shows the necessity of starting with a clearly defined 
mechanism, which can be precisely implemented (not phenomenologically invoked) 
and from which clear predictions can be systematically derived. The approach 
we have described, and the results we have obtained, in this Letter follow 
this philosophy closely and we expect that comparison of our results with 
experiments in the future will help to clarify the effect of molecular crowding 
and of resource depletion on diffusion.

\acknowledgments 

We thank Tommaso Biancalani, Andrea Gambassi and Gunter Sch\"{u}tz for useful discussions.

\end{document}